\documentclass[]{easychair}

\usepackage{todonotes}
\usepackage{paralist}

\usepackage[utf8]{inputenc} % allow utf-8 input
\usepackage[T1]{fontenc}    % use 8-bit T1 fonts
\usepackage{url}            % simple URL typesetting
\usepackage{booktabs}       % professional-quality tables
\usepackage{amsfonts}       % blackboard math symbols
\usepackage{nicefrac}       % compact symbols for 1/2, etc.
\usepackage{microtype}      % microtypography
\usepackage{xcolor}         % colors
\usepackage{comment}

\usepackage{booktabs} 
\usepackage{subcaption}
\usepackage{graphicx} % margin definition
\usepackage{pgfplots} % graphics
\pgfplotscreateplotcyclelist{rw}
{solid, mark repeat = 1, mark phase = 1, mark = *, black\\
	solid, mark repeat = 1, mark phase = 1, mark = square*, black\\}

\usepackage{tikz} % graphics
\usetikzlibrary{shapes,arrows}
\usepackage{pgfplots}
\usepackage{float}
\usepackage{bussproofs}
\usepackage{xspace}
\usepackage{amsmath}
\usepackage{mathtools}
\usepackage{doi}
\usepackage{multicol}
\usepackage{amsthm}
\newtheorem{definition}{Definition}
\usepackage{mathpartir}

\def\holfour{\textsf{HOL4}\xspace}

\def\vampire{\textsf{Vampire}\xspace}

\def\cvc{\textsf{CVC5}\xspace}
\def\zthree{\textsf{Z3}\xspace}
\def\sml{\textsf{SML}\xspace}

\def\sml{\textsf{Standard ML}\xspace}
\usepackage{color}

\usetikzlibrary{arrows,shapes,calc}
\usetikzlibrary{trees,positioning,fit}
\tikzstyle{arrow}=[draw,-to,thick]
\tikzstyle{embedding} = [draw, minimum width=8mm, minimum height=6mm]
\tikzstyle{nnop} = [draw, minimum width=8mm, minimum height=8mm, rounded 
corners]
\tikzstyle{block} =
[rectangle, draw,
minimum width=9mm, text centered, rounded corners, minimum height=2.0em]
\tikzstyle{smallblock} =
[rectangle, draw,
minimum width=9mm, text centered, minimum height=2.0em]

\tikzstyle{line}=[draw]
\tikzstyle{cloud} =
[draw, text centered, ellipse, minimum height=2.0em, minimum width=9em]
\usepackage[]{listings}

\newcommand{\mloop}{\mathit{loop}}
\newcommand{\mloopt}{\mathit{loop2}}
\newcommand{\mcompr}{\mathit{compr}}
\newcommand{\mcond}{\mathit{cond}}
\newcommand{\mmod}{\mathit{mod}}
\newcommand{\mathleft}{\@fleqntrue\@mathmargin0pt}
\newcommand{\ite}[3]{\text{if } #1 \text{ then } #2 \text{ else } #3}
\newcommand{\mdiv}{\mathit{div}}
\newcommand{\mmods}{\ \mathit{mod}\ }
\newcommand{\mdivs}{\ \mathit{div}\ }

\usepackage{enumitem}
\setlist[itemize]{nosep, topsep=0pt, partopsep=1pt, leftmargin=*, labelsep=5pt}

\title{A Mathematical Benchmark for \\ Inductive Theorem Provers}
\author{Thibault Gauthier, Chad E. Brown, Mikol\'{a}\v{s} Janota, Josef Urban}
\institute{Czech Technical University in Prague\\
	\email{email@thibaultgauthier.fr}, \email{mikolas.janota@gmail.com}, 
	\email{josef.urban@gmail.com}
}
\authorrunning{Gauthier et al.}
\titlerunning{A Mathematical Benchmark for Inductive Theorem Provers}
  
\begin{document}

\maketitle

\begin{abstract}
	We present a benchmark of 29687 problems derived from 
	the 
	On-Line Encyclopedia of 
	Integer Sequences (OEIS). Each problem expresses the equivalence of two 
	syntactically
	different programs generating the same OEIS sequence. Such programs were 
	conjectured by a learning-guided synthesis system using a 
	language with looping operators. The operators implement recursion, and 
	thus many of 
	the proofs require induction on natural numbers. 
	% The benchmark is based on OEIS problems from a wide area of mathematical 
	% domains 
	% and contains problems of varying difficulty.
	The benchmark contains problems of varying difficulty from a wide area of mathematical 
	domains. 
	We believe that these characteristics will make it an effective judge for 
	the 
	progress of inductive theorem provers in this domain for years to come.
%	\todo{We're not 100\% sure that they do generate the same sequence, I would 
%	say that this is how we actually define equivalence.}
\end{abstract}

\section{Introduction: Induction, OEIS and Related Work}

In mathematics, the principle of induction is an essential tool for proving 
various conjectures. This is especially true if the problem can be expressed as 
an arithmetical problem. Our goal in this project is to provide a benchmark to 
test the progress of theorem provers at proving mathematical formulas. Our hope 
is that in the long run, mathematicians will be able to use these tools to 
automatically prove non-trivial conjectures.

The 29,687 problems in our benchmark were derived automatically by running a 
program synthesis algorithm on the OEIS~\cite{oeis}. The OEIS repository 
archives common (and less common) integer sequences observed in combinatorics, 
group theory, geometry, etc. Every time two programs $P$ and $Q$ generate the 
same OEIS sequence, we can make the conjecture that $\forall x \in \mathbb{N}.\ 
f_P(x) = f_Q(x)$. Depending on the number of terms tested, this conjecture is 
more or less likely to be true. In this work, to minimize the number of false 
conjectures, only programs that cover all terms of a sequence available in the 
OEIS repository are considered. Equalities are also tested on additional inputs 
not present in the OEIS data.

We believe that this benchmark is a good challenge for inductive theorem 
provers as it is naturally grounded in mathematical theories. Our benchmark 
contains some easy problems, as shown in our first evaluation, but most of them 
are out of reach of the current best inductive theorem provers while being 
quite easy for a university student. We hope that our benchmarks will help 
developers bridge that gap. As a starting point for this research endeavor, we 
provide a translation of our benchmark to SMT-LIB~\cite{barrett2010smt} and 
evaluation baselines for future comparisons.

There already exist two benchmarks for inductive theorem provers. Compared to 
our benchmark, both of them are mostly focused on problems related to software 
verification. The first one is the ``Tons of Inductive Problems'' 
benchmark~\cite{DBLP:conf/mkm/ClaessenJRS15} including 340 problems about 
lists, natural numbers, binary trees, and integers originating from Isabelle, 
Agda and CLAM translated to SMT-LIB or to WhyML\@. The second set of problems is 
``Inductive Benchmarks for Automated 
Reasoning''~\cite{DBLP:conf/mkm/HajduHKSV21}. It consists of 3,516 problems 
about lists, natural numbers, trees, and integers. These problems were either 
handcrafted or inspired by software verification problems. To test the limit of 
inductive problems, multiple versions of the same problem with increased 
parameters were included. All those problems were translated to SMT-LIB and 
some of them to formats supported by Zipperposition and ACL2.

% and we provide translations to SMT-LIB and TH0.
%
%inductive provers (e.g., ACL2 [3], Zeno [15] or Imandra [11])
%tactic-based theorem provers~\cite{yutaka,tactician,tactictoe} re-using 
%induction 
%tactics from their respective proof assistants.
%smt-solvers~\cite{cvc4,z3}.
%higher-order theorem provers~\cite{satallax}
%
%Even standard first-order ATPs are now starting to support inductive 
%reasoning~\cite{10.1007/978-3-030-53518-6_8}.
%% with support for induction
%The TIP benchmarks were taken from problems in proof assistant such as 
%Isabelle/HOL or Agda.
%Our motivation for creating this benchmark is two-fold.
%One is to provide a way to evaluate improvement of the state-of-the-art
% in inductive theorem proving for problems requiring induction.
%The first one is to create a large dataset of inductive problems containing 
%problems of various difficulties. Where as previous problems were targeting 
%formulazition about computer science such as grammar and algebraic datatypes,
%we only focus on integers 
%provided by far the largest benchmark.
%This benchmark 

\section{Programming Language}\label{sec:programs}
We now present the programming language used in our benchmark.
This language contains the same operators as 
our system for synthesizing programs from integers 
sequences~\cite{oeis-synthesis-arxiv}. There, a simplified presentation 
of the semantic of the operators was given. In this paper, we present a formal 
version that matches the definitions given in the SMT problems.
%Our domain-specific
%language relies on only several basic arithmetical constants and operators 
%such 
%as $0, +, \mdiv$, and fundamental programming 
%constructs such as variables and loops. 

\paragraph{Syntax}
The set $\mathbb{P}$ of programs in our language is inductively defined to be 
the 
smallest set such that
$0,1,2,X,Y\in \mathbb{P}$, and if $A,B,C,F,G\in \mathbb{P}$ then $A + B$, 
$A - B$, $A \times B$, $A \mdivs B$, $A \mmods B$,$\mcond(A,B,C)$, 
$\mloop(F,A,B)$, $\mloopt (F,G,A,B,C)$, $\mcompr(F,A) \in \mathbb{P}$.

In the rest of this paper, we refer to $\mloop$, $\mloopt$, and $\mcompr$ as 
\textit{looping operators}, and we refer to the other operators ($0$, $1$, $2$, 
$X$, 
$Y$, $+$, $-$, $\times$, $\mdiv$, $\mmod$, and $\mcond$) as \textit{first-order 
operators}. 
The first argument of $\mloop,\mcompr$ 
and the first two arguments of $\mloopt$ are called higher-order arguments 
(designated by $F$ and $G$ in the previous definition).
If a variable ($X$ or $Y$) appears in a higher-order argument it is said to 
be bounded otherwise it is said to be free. 
We say that a program $P$ syntactically depends on a variable if this variable 
appears free in $P$. 
The symbols $0,1,2,+,-,\times,\mdiv,\mmod$ are 
overloaded and may refer to program operators, 
\sml functions or SMT functions depending on the context.

\paragraph{Semantics}
Each program $P$ is interpreted by a function $f_P: (x,y) \in \mathbb{Z}^2
\mapsto f_P(x,y) \in \mathbb{Z}$.  The interpretation $f_P$
is recursively defined for every program $P$ by:

\begin{align*}
&f_{0} (x,y) := 0,\ f_{1} (x,y) := 1,\ f_X(x,y) = x,\ f_Y(x,y)=y\\
&f_{A + B} (x,y) := f_A(x,y) + f_B(x,y),\ f_{A - B}(x,y) := f_A(x,y) - 
f_B(x,y)\\
&f_{A \times B} (x,y) := f_A(x,y) \times f_B(x,y),\ f_{A \mdiv B}(x,y) := 
f_A(x,y) \mdivs f_B(x,y)\\
&f_{A \mmod B}(x,y) := f_A(x,y) \mmods f_B(x,y)\\
&f_{\mcond(A,B,C)}(x,y) :=\ \ite{f_A(x,y) \leq 0}{f_b(x,y)}{f_c(x,y)}\\
&f_{\mloop(F,A,B)}(x,y) = u(f_A(x,y),f_B(x,y))\\
&\hspace{24mm} \mbox{where } u(x,y) = \ite{x \leq 0}{y}{f_F(u(x-1,y),x)}\\
&f_{\mloopt(F,G,A,B,C)}(x,y) := u(f_A(x,y),f_B(x,y),f_C(x,y))\\
&\hspace{24mm} \mbox{where } u(x,y,z) = \ite{x \leq 
0}{y}{f_F(u(x-1,y,z),v(x-1,y,z))}\\
&\hspace{24mm} \mbox{and } v(x,y,z) = \ite{x \leq 
0}{z}{f_G(u(x-1,y,z),v(x-1,y,z))}\\
&f_{\mcompr(F,A)}(x,y) := u(f_A(x,y))\\
&\hspace{24mm} \mbox{where } t(x) = \ite{f_F(x,0) \leq 0}{x}{t(x+1)}\\
&\hspace{24mm} \mbox{and } u(x) = \ite{x \leq 0}{t(0)}{t(u(x-1) + 1)}
\end{align*}

The constants and functions used (outside 
program indices) in this recursive definition.
follow the semantics of \sml~\cite{harper1986standard}. Note that 
the functions created from $\mmod,\mdiv,\mcompr$ may not be total.

We now give an intuition for the semantics
of the looping operators. In this informal description, we do not show the 
trivial behavior of the following auxiliary sequences on negative indices.
The operator $\mloop$ is constructing a recursive sequence $u_n$ and returns 
the value $u_{f_A(x,y)}$. This sequence is defined by:
\[u_0 = f_B(x,y), u_n = f_F(u_{n-1},n)\]

The operator $\mloopt$ is constructing two mutually recursive sequences 
$u_n$ and $v_n$. It returns the value $u_{f_A(x,y)}$. These sequences are 
defined by:
\[u_0 = f_B(x,y), v_0 = f_C(x,y), u_n = f_F(u_{n-1},v_{n-1}), v_n = 
f_G(u_{n-1},v_{n-1})\]

The operator $\mcompr$ constructs a sequence $u_n$ and returns the value 
$u_{f_A(x,y)}$. The sequence $u_n$ returns the $(n+1)^{\mathit{th}}$ smallest 
non-negative 
integer $x$ satisfying $f_F(x,0) \leq 0$. The auxiliary function $t(x)$ 
searches for the next number $y \geq x$ satisfying $f_F(y,0) \leq 0$.

%indcution petra
\paragraph{Execution}
When executing programs we limit the number of steps to 
100,000 abstract time units per
call during the self-learning experiment (Section~\ref{sec:sl}) and during the 
cyclicity checks (Section~\ref{sec:induct}) 
This time limit is increased to 1,000,000 abstract time units per call during 
equality verification (Section~\ref{sec:pb}). 
The number of abstract time units consumed by the execution of a program is an 
estimate proportional to the number of CPU instructions needed for each 
operator. It is 5 for $\mmod$ and $\mdiv$. It is 1 for all other first-order 
operators. In case the absolute value of the integer returned by the operator 
is bigger than $2^{64}$, the number of digits in this integer is used as an 
estimate.
When generating a sequence, the timeout for the current call is increased by 
adding the unused time from the previous calls.
On top of this, the execution stops and fails when a number with absolute value 
greater than $10^{285}$ is produced.

\paragraph{Properties}
The \textbf{size} of a program is measured by counting with repetition the 
number of operators 
composing it. 
The \textbf{speed} of a program is measured by the total number of abstract 
time units used when generating a sequence.
If two programs have the same size (respectively speed) a fixed 
total order is used to determine which one is the smallest (respectively 
fastest). 
\begin{definition}[Cover]
We say that a program \textbf{covers} (or \textbf{is a solution 
for}) an OEIS sequence $(s_x)_{0\leq x\leq n}$ if and only if $\forall x\in 
\mathbb{Z}.\ 0\leq x\leq n \Rightarrow f_P(x,0) = s_x$.
\end{definition}

\section{Benchmark}\label{sec:benchmark}
Our benchmark consists of problems of the form $\forall x \in \mathbb{N}.\  
f_\textit{Small}(x) = f_\textit{Fast}(x)$ where $f_\textit{Small}$ and 
$f_\textit{Fast}$ are functions created from the small program 
$\textit{Small}$ 
and a fast program $\textit{Fast}$. During the checking phase of the 
self-learning loop, we only test and select a program if it does not depend on 
$Y$ (its higher-order arguments may depend on $Y$). 
This way, we are 
able to express $f_\textit{Small}$ and $f_\textit{Fast}$ as unary functions.
An explanation of how these functions are defined in our SMT problems is given 
in Section~\ref{sec:translation}.

\subsection{Short Overview of the Self-Learning System}\label{sec:sl}
\begin{figure}[t]
	\centering
	\begin{tikzpicture}[scale=1.0]
	\begin{axis}[
	legend style={anchor=north east, at={(0.95,0.4)}},
	width=\textwidth,
	height=0.25*\textwidth,xmin=-0.2, xmax=209.2,
	ymin=0, ymax=50000,
	cycle list name=rw,
	xtick = {0,50,100,150,200},
	scaled y ticks = false
	]
	\addplot table[x=gen, y=sol] {gensol};
	\end{axis}
	\end{tikzpicture}
	\caption{OEIS sequences covered after $x$-th generation\label{fig:gensol}}
\end{figure}
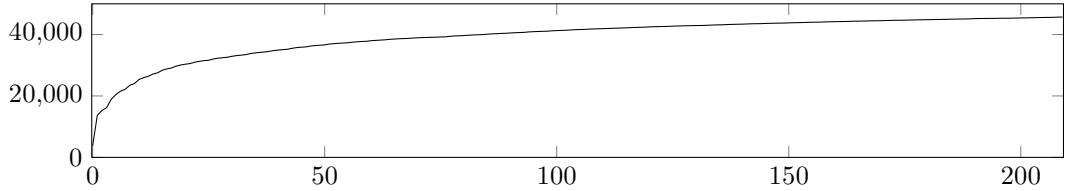

The programs present in the benchmark were discovered through self-learning. The 
system gradually discovers on its own programs for OEIS sequences. The 
self-learning loop was run for 209 generations instead of 25 generations in~\cite{oeis-synthesis-arxiv}.
At each generation, we recorded the smallest and fastest programs 
discovered so far for each OEIS sequence (instead of only the smallest as in~\cite{oeis-synthesis-arxiv}).
Each generation consists of a synthesis phase, a checking phase and a learning 
phase. During the synthesis phase, programs are created using a probability 
distribution on operators given a target sequence returned by the learning 
phase. At generation 0, a 
random probability distribution is used. During the checking phase, we 
check that the programs created cover the target sequence or any other OEIS 
sequence. During the learning phase, we train a tree neural network~\cite{DBLP:conf/mkm/Gauthier20}
to predict given a target sequence the smallest and the fastest programs 
discovered so far generating it.
This process repeats finding solutions for more and more OEIS sequences as 
depicted in Figure~\ref{fig:gensol}.

\pgfplotscreateplotcyclelist{rw}
{solid, mark repeat = 100, mark phase = 500, mark = *, black\\
	solid, mark repeat = 100, mark phase = 1000, mark = square*, black\\}

%Change in the self-learning
%the self-learning loop was run for 209 generations 

\subsection{Problems in the Benchmark}\label{sec:pb}

At the end of the self-learning loop, a program covering 
an OEIS sequence is found for 45691 sequences. 
This number is reduced to 34171 when looking for sequences where the smallest 
program $\mathit{Small}$ and the fastest program $\mathit{Fast}$ are 
syntactically different.
From each pair of programs, we construct the problem $\forall x\in 
\mathbb{N}.\ f_{\mathit{Small}}(x) = f_{\mathit{Fast}}(x)$. 
After regrouping the sequences that generate the same exact equations (this may 
happen because some OEIS sequences are prefixes of other OEIS sequences), we get 
32124 unique problems.
A typical OEIS sequence may contain anywhere from a few terms to about 200 
terms with the most common number of terms being between 20 and 50. 
For OEIS sequences with less than 100 terms, we further check that both 
programs are indeed equal on the first 100 non-negative integers with a given 
time limit. If, when computing, one of the programs exceeds the execution 
limits,
 the equality is still considered potentially correct if all equality checks 
pass before an error is raised. Indeed, some interesting equalities occur
between long-running programs or fast-increasing programs.
After this last check, our dataset is reduced 29687 problems. These problems, 
after translation to SMT-LIB, constitute the released benchmark available at 
\url{grid01.ciirc.cvut.cz/~thibault/oeis-smt.tar.gz}. The code for running the 
self-learning loop and translating 
the problems to SMT-LIB is available at 
\url{https://github.com/barakeel/oeis-synthesis}. After running the 
following commands in an interactive \holfour session 
will produce the SMT benchmark from the discovered solutions stored in the file
\texttt{model/itsol209}
in the subdirectory \texttt{oeis-smt}:
\begin{verbatim}
load "smt"; smt.export_smt2 true "oeis-smt" "model/itsol209";
\end{verbatim}
5435 problems (included in the benchmark) could not be fully verified because 
of the execution limits.
The non-verified problems are listed in the file \textit{all\_nonverified100}
for further analysis.

\subsection{Examples}\label{sec:examples}
Here are a few examples of the benchmark problems derived from famous (and less 
famous) sequences in the 
OEIS\@. For each of those problems, we first give the OEIS sequence number 
and its description and interpret the meaning of the 
derived equality between the two programs. In this list, the 
equality $\mathit{Small}=\mathit{Fast}$ is used as a shorthand for the 
conjecture $\forall 
x\in \mathbb{N}.\ f_{\mathit{Small}}(x) = f_{\mathit{Fast}}(x)$

\begin{itemize}
\item A217, triangular numbers:
 \[\mloop(X + Y,X,0) = ((X \times X) + X) \mdivs 2\]
In this example, a loop is used to computes the sum of the first $n$ 
non-negative integers, thus 
the equation can be rewritten in mathematical form as 
$\sum_{i=0}^n i = \frac{n \times n + n}{2}$.

\item A537, sum of first n cubes:
  \[\mloop ((Y \times Y) \times  Y + X, X, 0) = 
    \mloop (X \times X, 1, ((X \times X) + X) \mdivs 2)\]
The loop on the right-hand side has a bound of 1. So it simply applies  
the squaring function $X \times X$ once to the initial value of the loop. In 
mathematical notation, this conjecture can be expressed as 
$\sum_{i=0}^n i^3 = (\frac{n \times n + n}{2})^2$

\item A45, Fibonacci numbers:
\[\mloopt (X + Y, X, X, 0, 1) = \mcond (X,0,\mloopt (X + Y, X, X - 2, 1, 1))\]
On the left-hand side is the expected definition for Fibonacci numbers. On the 
right-hand side the fast program seems to be saving some computation by starting 
the loop two steps later with higher initial values.
Due to the similarity between the loops in the two programs, this problem may 
be proven without induction by unrolling the % first
loop twice.
However, a proof using induction might be easier to find.

\item A79, powers of 2:
\[\mloop (X + X, X, 1) = 
  \mloop (X + X, X \mmods 2, 
  \mloop (X \times X, 1, \mloop (X + X, X \mdivs 2, 1)))\]
 
Two bounded loops are used to compose functions in the right-hand side of this 
equation. There, the result of $\mloop (X + X, X \mdivs 2, 1))$ is squared and 
then multiplied $x \mmods 2$ timesby $2$. This
conjecture can thus be rewritten as $2 ^ x = 2 ^ {(x \mmods 2)} \times (2^{(x 
\mdivs 
2)})^2$. The proof will likely require inductive reasoning to prove the lemma 
$2^x 
\times 2^y = 2^{x+y}$. The fast program uses the first step of the fast 
exponentiation algorithm to speed up the computation.

\item A165, double factorial of even numbers, $(2n)!! = 2^n \times n!$: 
\[\mloop(2 \times (X \times Y), X, 1) =
  \mloop(X + X, X, 1) \times \mloop(X \times Y, X, 1)\]

The double factorial of $2n$ is by definition $(2n)!! = \prod_{k=1}^n 2k$
Thus coincidentally, this equation gives an implementation on each side of the 
equation of the two formulas given in the OEIS.  A proof of 
this statement is expected to require inductive reasoning. 
%This 
%statement contains 
%fast-increasing functions and therefore could not be verified 
%through testing with the given resource constraints.

%.

%The OEIS sequence contains only 33 terms. This may cause 
%the full fast exponentiation algorithm to be slower than this simple 
%optimization.

\end{itemize}

\subsection{Problems Requiring Induction}\label{sec:induct}
One motivation\footnote{There are multiple motivations for this benchmark, some 
of them being very pragmatic. Our
  OEIS program synthesis systems produce thousands of more and more complex 
  programs that may look quite 
  \emph{alien}~\cite{DBLP:journals/corr/abs-2301-11479}. It may take a 
  nontrivial amount of time to decide if such programs are correct and human 
  mathematicians do not scale to the number of such problems we are currently 
  generating.}
for this benchmark is to test provers on mathematical 
problems requiring one or multiple inductions. 
Since our programs consist of looping constructs, we believe that is the case 
for the majority of the problems in our benchmark. However, some 
of the problems can be solved without induction. An easily recognizable case is 
when the programs $\mathit{Small}$ and $\mathit{Fast}$ do not contain any loop 
or when all their loops are bounded by a constant. Such problems may occur in 
our dataset. Therefore, in the following, we design \emph{syntactic and semantic 
tests} to detect if an equality contains at least one ``proper'' top-level loop.

Given an equality $\forall x\in \mathbb{N}.\ f_{\mathit{Small}}(x) = 
f_{\mathit{Fast}}(x)$, we select a looping 
subprogram (a subprogram whose root operator is a looping operator) in 
$\mathit{Small}$ and 
$\mathit{Fast}$ if it appears at a position that is not under another looping 
subprogram and 
if its exact formulation does not appear more than once in the equational 
problem.
We will say that a problem \emph{passes a test} if there exists at least 
one such top-level loop which satisfies this test. From the syntactic and 
semantic tests, we carve 
out two subsets of the released benchmark. Problems that pass all syntactic 
tests are listed in the file \texttt{aind\_syn} and problems that pass all 
syntactic and semantic tests are listed in the file \texttt{aind\_sem}.
In general, it is a hard problem to determine if a problem will require 
induction a priori, and the following tests are trying to achieve a trade-off
between ruling out problems that do not require 
induction and keeping problems that do.

\paragraph{Syntactic Tests}
The following syntactic tests are performed on the looping subprograms 
of the form $\mloop(F_1,A_1,B_1), \mloopt(F_2,G_2,A_2,B_2,C_2), 
\mcompr(F_3,A_3)$:
\begin{itemize}
\item The bounds $A_1,A_2,A_3$ and the subprogram $F_1$ must depend on $X$.
\item Either $F_2$ or $G_2$ must depend on $X$ and $Y$.
\end{itemize}

\paragraph{Semantic Tests}
The semantic tests will try to detect cases where a prover does not require 
induction even though the problem passes the syntactic tests.
For instance, top-level looping subprograms may use $X \mmods 3$ or $2 - X$ as a 
bound or some behavior in $F_1,F_2$ or $G_2$ may
result in a proof that unrolls the loop a finite amount of time.
Thus, instead of testing for syntactic dependency on a 
variable, we will run the subprograms and test for acyclicity in their output.

\begin{definition}[Acyclicity of an integer sequence -- tailored to our setting]
We say that a finite sequence of integers $a_0,a_1,\ldots a_{39}$ 
is \emph{acyclic} if and only if the sequence $a_9,\ldots,a_{39}$ does not contain a 
cycle with a period ranging from $1$ to $15$.
\end{definition}

\begin{definition}[Acyclicity of a program]
A program $P$ is \emph{acyclic} on $x$ if and only if:
$\forall y\in \mathbb{Z}.\ 0 \leq y \leq 9 \Rightarrow (f_P(x,y))_{0 \leq x 
	\leq 39} \mbox { is acyclic}$.\\ 
A program $P$ is \emph{acyclic} on $y$ if and only if:
$\forall x\in \mathbb{Z}.\ 0 \leq x \leq 9 \Rightarrow (f_P(x,y))_{0 \leq y 
\leq 39} \mbox { is acyclic}$.\\
In practice, we chose to make the test fail if one of the sequence cannot be 
produced because of the execution limits. In such situations, the program $P$ will not be 
considered acyclic.
\end{definition}

The following semantic tests are performed on the top-level looping subprograms 
of the form $\mloop(F_1,A_1,B_1), \mloopt(F_2,G_2,A_2,B_2,C_2), 
\mcompr(F_3,A_3)$:
\begin{itemize}
\item The bounds $A_1,A_2,A_3$ and the subprogram $F_1$ must be 
acyclic on $x$. \\When checking for cyclicity in the bounds, all 
negative program outputs are mapped to 0 before checking for cycles.
\item Either $F_2$ or $G_2$ must be acyclic on $x$ and acyclic on $y$.
\item The looping subprogram itself must be acyclic on $x$.
\end{itemize}

\section{Translation to SMT-LIB}\label{sec:translation}
We now translate the 29687 problems in our benchmark (see 
Section~\ref{sec:benchmark}) to SMT-LIB\@. 
These SMT problems consist of definitions for $f_\textit{Small}$ and 
$f_\textit{Fast}$, and the negated conjecture:
\[\exists c.\ c \geq 0 \wedge \neg (f_\textit{Small}(c) = f_\textit{Fast}(c))\]
  
Instantiating the semantic definitions given in Section~\ref{sec:programs}, we 
can 
recursively make definitions for each subprogram in the two 
top-level programs. In these definitions, the \sml functions are replaced by 
their SMT counterparts.
This process creates a new definition for each subprogram. In order to simplify 
the SMT problem we expand definitions for first-order operators.
We also minimize the number of arguments of each 
function. Indeed, if a program $P$ does not depend on $Y$ (respectively $X$,$X$ 
and $Y$), we can define a function $f_P$ with one argument such that
$f_P(x):=f_P(x,y)$ (respectively $f_P(y):=f_P(x,y)$, $f_P():= f_P(x,y)$).
We illustrate the process of creating SMT definitions on the program 
$\textit{Small}=\mloop(X + Y,X,0)$:
\begin{align*}
&f_1 () = 1, f_X (x) = x, f_{X + Y}(x,y) := x + y\\
&u_{\mloop(X + Y,X,1)}(x,y) = \ite{x \leq 0}{y}{f_{X + 
Y}(u(x-1,y),x)}\\
&f_{\mloop(X + Y,X,1)}(x) = u(f_X(x),f_1())
\end{align*}

The third line is derived by expanding definitions of first-order operators 
until a looping subprogram is reached. 
In our example, we have $f_{X + Y}(x,y) = f_{X}(x,y) + f_{Y}(x,y) = x + y$.
In a more general example where a looping subprogram $Q$ appears under the 
first-order part, we would for instance get $f_{2 \times X + Q}(x,y) = 2 \times 
x + f_{Q}(x,y)$.
The in-lining of expanded definitions inside definitions for loops is left to 
the provers.
In the released SMT problems, program indices in the definitions 
are replaced by integer indices.

\paragraph{Totality of the Functions}
Some of our \sml functions may initially be partial because of the operators 
$\mdiv,\mmod,\mcompr$. These are translated to total SMT functions.
Therefore, a proof of the equality between two functions $f_\textit{Small}$,
$f_\textit{Fast}$ with 
 respective initial domain $D_\textit{Small}$,
$D_\textit{Fast}$ is only a proof that they are equal on 
$D_\textit{Fast} \cap D_\textit{Small}$. Functions from our problems are 
expected to be total. This is especially the case if we consider only those 
that fully passed 
the verification test on the first 100 non-negative integers.

\paragraph{Towards More General Conjectures}
In our benchmark, all of the conjectures are of the form 
$\forall x. x \geq 0 \Rightarrow f_P(x) = f_Q(x)$. In particular, we cannot 
express conjectures that quantifies over $P$ or $Q$. To solve this issue, 
one can use an evaluation function $\mathit{eval}(P,x,y):=f_P(x,y)$ that makes 
$P$ a 
proper argument instead of an index. This alternative translation requires to 
declare the syntactic constructs as SMT functions on a SMT sort for programs 
and replace in the defining equations (semantics) for each operator every 
instance of $f_P(x,y)$ by $eval(P,x,y)$ to create the SMT axioms. After this 
transformation, we can, for example, express the conjecture of finding an 
increasing function as $\exists P.\ \mathit{eval}(P,x,0) \leq 
\mathit{eval}(P,x+1,0)$. We 
did not 
use this $eval$ 
encoding in our benchmark since we do not need it for our current conjectures,
% as they did not require such encoding
and also because we 
observed that it makes the problems more difficult for the provers.

\section{Experiments}\label{sec:experiment}
% Detailed results for vampire z3 and cvc5
% Comments on how the results change if we remove problems that do not require 
% induction.

In the following experiments, we test the performance of three state-of-the-art 
provers \vampire~\cite{DBLP:conf/cav/KovacsV13},
\cvc~\cite{DBLP:conf/cade/KremerRBT22,DBLP:conf/tacas/BarbosaBBKLMMMN22}
and 
\zthree~\cite{demoura:2008:zes:1792734.1792766} on our benchmark. 
\vampire is run with an induction
schedule~\cite{DBLP:conf/cade/HozzovaKV21} suggested by its developers;
\cvc is run with its induction flag on~\cite{reynolds-vmcai15}. The addition of support for 
arithmetical induction 
is recent in these two % previous
provers.
The prover \zthree has not yet been given such support and 
therefore can only solve problems that 
do not require induction. 
All provers are run on all the problems with a timeout 
of 60 seconds for each problem.

Table~\ref{tab:res1} shows the results of running \zthree, \vampire and 
\cvc on the benchmark.
For \cvc we also show the results after strengthening the conjecture to include 
equality of additional terms.
For example, C1 adds equality on the successor, C2 also on the successor of successor, etc.
C2x changes the conjecture to equality on $2x$ and $2x + 1$. 
These additional methods have a significant influence on \cvc, but not on 
\zthree and \vampire.
We have also tried to change the conjecture
to strong induction (equality on all previous numbers). However, it has 
practically no effect on \cvc.
For each method, we also show the results on the 23163 problems that pass the syntactic filtering (Section~\ref{sec:induct}), results on
  the 16197 problems that pass the semantic filtering (Section~\ref{sec:induct}), and 
  results on the 5435 problems where the extended verification
  (testing on 100 terms - Section~\ref{sec:pb}) fails.

  The fact that \zthree solves only 7 problems after the strongest semantic
  filtering suggests that induction is very likely to be needed on the
  16197 semantically filtered problems. In total, we can prove 2686 of
  those problems, which is 16.58\% . An interesting result is the 212
  problems (3.90\% of the 5435) where we can prove equality of the
  functions, but our extended equality verification (testing on 100 terms)
  procedure fails on them. Most often this means that these are 
  fast-growing functions where the normal computation of the numerical
  values overflows on larger inputs. While 3.90\% is not much, it
  demonstrates a real value added by automated reasoning compared to
  just running extended testing. The joint performance of all systems
  on the full benchmark is 26.67\% (8215 out of 29687) and the
  performance on the syntactically filtered problems is 16.16\% (3743
  out of 23163). \zthree is the best system on the full benchmark, while \cvc
  performs best on the filtered problems where induction is likely often
  needed. \cvc is also quite orthogonal to \zthree and \vampire, adding many 
  solutions to both.

% (/ 3743 23163.0) = 0.16159
% (/ 8215 29687.0) = 0.2767
% (/ 2686 16197.0) = 0.1658
% (/ 212 5435.0) = 0.0390

%grep A217.smt2 *
%grep A537.smt2 *
%grep A45.smt2 *
%grep A79.smt2 *
From the five examples presented in Section~\ref{sec:examples}, the 
problem derived from the triangular numbers is 
solved by \cvc. Strengthening the conjecture (to the successor - method C1) can also solve the problem
induced by the Fibonacci numbers.
\vampire is able to solve the problem about double factorials and \zthree can 
not solve any of them. 
\vampire can also solve the triangular numbers problem when using its default 
induction schedule instead of the suggested one.

% Using the default 
% induction schedule for \vampire instead of the suggested one, we found out that 
% it is also able to prove the problem about triangular numbers.
% Vampire should also be able to solve some of them them.

\begin{table}[t]
	\small
  \begin{tabular}{llllllllllllll}\toprule
System & Z & V & C &          C1 & C2 &        C3 & C4  
&       
C5 & C6 & C8 & C2x &  All\\        % Cs\\
NoFilt & 4757 & 2195 & 2428 & 3793 & 4030 & 4100 & 4084 & 3962 & 3796 & 3451 & 3557  & 8215 \\ % 2418 \\
    SynFilt &487 & 278 & 893 & 2258 & 2547 & 2701 & 2699 & 2590 & 2447 & 2156 & 2015 &  3743 \\  % 884  \\
    SemFilt &7 & 83 & 504 & 1799 & 2059 & 2235 & 2240 & 2146 & 2021 & 1786 & 1501 &    2686\\    %498 \\
    NonVer & 2 & 97 & 21 & 22 & 48 & 33 & 76 & 39 & 29 & 8 & 5 & 212 \\\bottomrule %21 
\end{tabular}
\caption{Problems solved by each of the methods. NoFilt means no
  filtering (results on the whole benchmark), SynFilt are results on
  the 23163 problems that pass the syntactic filtering, SemFilt are results on
  the 16197 problems that pass the semantic filtering, and NonVer are
  results on the 5435 problems where the extended verification
  (testing) fails.}\label{tab:res1}
\end{table}

%TODO: refactor, possibly appendix/web?

\section{Conclusion}
In this work, we relied on a self-learning system to create small and fast 
programs for each OEIS sequence. Subsequently, we created a benchmark of 
29,687 SMT problems, asserting that the two programs produce identical 
sequences. We then asked automated theorem provers and SMT solvers to solve
these problems and analyzed the results. Ultimately, we discovered a simple 
method to enhance the performance of \cvc on inductive problems.

In the future, we aim to use our benchmark to reveal the limitations of 
inductive 
theorem provers available in proof assistants and explore ways to improve them. 
This will enable us to assess the impact of various techniques developed over 
the years, such as term synthesis~\cite{hipspec13}, 
rippling~\cite{DBLP:conf/birthday/JohanssonDB10}, and 
template-based conjecturing~\cite{DBLP:journals/corr/abs-2212-11151}. To 
achieve this, we 
will need to translate the problems into the specific format required by each 
of these inductive theorem provers.
Last but not least, we aim to explore the achieved results in the context of
program equivalence~\cite{DBLP:journals/stvr/GodlinS13}.

\bibliographystyle{plain}
\bibliography{biblio}

\appendix
\section{Further Examples of Programs and their Encoding}

% The following two programs for A257594\footnote{\url{oeis.org/A257594}} include division by 0, making their evaluation-based equality checking fail.
% Since division is total in SMT, both \vampire and \zthree can prove the corresponding SMT problem.

% \begin{small}
%   \begin{verbatim}
% ;; sequence(s): A257594
% ;; terms: 0 0 0 0 0 0 1 1 2 3 4 5 7 8 10
% ;; small program: (x * x) div ((2 * (loop (x * x) 2 2)) - x)
% ;; fast program: (x * x) div ((2 * (2 * (2 * (2 + 2)))) - x)
% (assert (forall ((x Int)) (= (f0 x) (* x x))))
% (assert (= g0 2))
% (assert (= h0 2))
% (assert (forall ((x Int) (y Int)) (= (u0 x y) (ite (<= x 0) y (f0 (u0 (- x 1) y))))))
% (assert (= v0 (u0 g0 h0)))
% (assert (forall ((x Int)) (= (small x) (div (* x x) (- (* 2 v0) x)))))
% (assert (forall ((x Int)) (= (fast x) (div (* x x) (- (* 2 (* 2 (* 2 (+ 2 2)))) x)))))
% (assert (exists ((c Int)) (and (>= c 0) (not (= (small c) (fast c))))))
% \end{verbatim}
% \end{small}

Here we show the SMT encoding of the problem created from A000165 -- double factorial of even numbers\footnote{\url{oeis.org/A000165}}.
The numbers grow too fast, making their evaluation-based equality checking fail. \vampire proves the problem and uses induction in its proof.

\begin{small}
\begin{verbatim}
;; sequence(s): A165
;; terms: 1 2 8 48 384 3840 46080 645120 10321920 185794560 3715891200 
;; 81749606400 1961990553600 51011754393600 1428329123020800 42849873690624000 
;; 1371195958099968000 46620662575398912000 1678343852714360832000 63777066403145711616000
;; small program: loop(2 * (x * y), x, 1)
;; fast program: loop(x + x, x, 1) * loop(x * y, x, 1)
(assert (forall ((x Int) (y Int)) (= (f0 x y) (* 2 (* x y)))))
(assert (forall ((x Int)) (= (g0 x) x)))
(assert (= h0 1))
(assert (forall ((x Int) (y Int)) (= (u0 x y) (ite (<= x 0) y (f0 (u0 (- x 1) y) x)))))
(assert (forall ((x Int)) (= (v0 x) (u0 (g0 x) h0))))
(assert (forall ((x Int)) (= (small x) (v0 x))))
(assert (forall ((x Int)) (= (f1 x) (+ x x))))
(assert (forall ((x Int)) (= (g1 x) x)))
(assert (= h1 1))
(assert (forall ((x Int) (y Int)) (= (u1 x y) (ite (<= x 0) y (f1 (u1 (- x 1) y))))))
(assert (forall ((x Int)) (= (v1 x) (u1 (g1 x) h1))))
(assert (forall ((x Int) (y Int)) (= (f2 x y) (* x y))))
(assert (forall ((x Int)) (= (g2 x) x)))
(assert (= h2 1))
(assert (forall ((x Int) (y Int)) (= (u2 x y) (ite (<= x 0) y (f2 (u2 (- x 1) y) x)))))
(assert (forall ((x Int)) (= (v2 x) (u2 (g2 x) h2))))
(assert (forall ((x Int)) (= (fast x) (* (v1 x) (v2 x)))))
(assert (exists ((c Int)) (and (>= c 0) (not (= (small c) (fast c))))))
\end{verbatim}
\end{small}

Next, we show the SMT encoding of the problem created from A45 -- the Fibonacci sequence\footnote{\url{oeis.org/A45}}.
\cvc proves the problem after strengthening the conjecture to the successor as follows:

\begin{small}
\begin{verbatim}
;; sequence(s): A45-A77373
;; terms: 0 1 1 2 3 5 8 13 21 34 55 89 144 233 377 610 987 1597 2584 4181
;; small program: loop2(x + y, x, x, 0, 1)
;; fast program: if x <= 0 then 0 else loop2(x + y, x, x - 2, 1, 1)
(assert (forall ((x Int) (y Int)) (= (f0 x y) (+ x y))))
(assert (forall ((x Int)) (= (g0 x) x)))
(assert (forall ((x Int)) (= (h0 x) x)))
(assert (= i0 0))
(assert (= j0 1))
(assert (forall ((x Int) (y Int) (z Int)) (= (u0 x y z) 
                 (ite (<= x 0) y (f0 (u0 (- x 1) y z) (v0 (- x 1) y z))))))
(assert (forall ((x Int) (y Int) (z Int)) (= (v0 x y z) 
                 (ite (<= x 0) z (g0 (u0 (- x 1) y z))))))
(assert (forall ((x Int)) (= (w0 x) (u0 (h0 x) i0 j0))))
(assert (forall ((x Int)) (= (small x) (w0 x))))
(assert (forall ((x Int) (y Int)) (= (f1 x y) (+ x y))))
(assert (forall ((x Int)) (= (g1 x) x)))
(assert (forall ((x Int)) (= (h1 x) (- x 2))))
(assert (= i1 1))
(assert (= j1 1))
(assert (forall ((x Int) (y Int) (z Int)) (= (u1 x y z) 
                 (ite (<= x 0) y (f1 (u1 (- x 1) y z) (v1 (- x 1) y z))))))
(assert (forall ((x Int) (y Int) (z Int)) (= (v1 x y z) 
                 (ite (<= x 0) z (g1 (u1 (- x 1) y z))))))
(assert (forall ((x Int)) (= (w1 x) (u1 (h1 x) i1 j1))))
(assert (forall ((x Int)) (= (fast x) (ite (<= x 0) 0 (w1 x)))))
(assert (exists ((c Int)) (and (>= c 0) (or (not (= (small (+ c 1)) (fast (+ c 1)))) 
                               (not (= (small c) (fast c))    )))))
\end{verbatim}
\end{small}

Finally, we show the encoding of  A180713\footnote{\url{oeis.org/A180713}} proved by \cvc only after changing the
conjecture to equality on $2x$ and $2x + 1$:
\begin{small}
\begin{verbatim}
;; sequence(s): A180713
;; terms: 0 4 6 11 12 16 18 23 24 28 30 35 36 40 42 47 48 52 54 59
;; small program: ((((((x div 2) * x) mod 2) + (x mod 2)) + x) + x) + x
;; fast program: (loop(loop(1, 2 - (x mod (2 + 2)), 2) + x, x mod 2, x) + x) + x
(assert (forall ((x Int)) (= (small x) 
                (+ (+ (+ (+ (mod (* (div x 2) x) 2) (mod x 2)) x) x) x))))
(assert (= f1 1))
(assert (forall ((x Int)) (= (g1 x) (- 2 (mod x (+ 2 2))))))
(assert (= h1 2))
(assert (forall ((x Int) (y Int)) (= (u1 x y) (ite (<= x 0) y f1))))
(assert (forall ((x Int)) (= (v1 x) (u1 (g1 x) h1))))
(assert (forall ((x Int)) (= (f0 x) (+ (v1 x) x))))
(assert (forall ((x Int)) (= (g0 x) (mod x 2))))
(assert (forall ((x Int)) (= (h0 x) x)))
(assert (forall ((x Int) (y Int)) (= (u0 x y) (ite (<= x 0) y (f0 (u0 (- x 1) y))))))
(assert (forall ((x Int)) (= (v0 x) (u0 (g0 x) (h0 x)))))
(assert (forall ((x Int)) (= (fast x) (+ (+ (v0 x) x) x))))
(assert (exists ((c Int)) (and (>= c 0) (or (not (= (small (* c 2)) (fast (* c 2)))) 
(not (= (small (* 2 (+ c 1))) (fast (* 2 (+ c 1))))) ))))
\end{verbatim}
\end{small}

\end{document}